# Enhancing Resilience of Power Systems against Typhoon Threats: A Hybrid Data-Model Driven Approach

Yang Li [1,*]

[1] School of Electrical Engineering, Northeast Electric Power University, Jilin, 132012, China

* Corresponding author. E-mail address: liyang@neepu.edu.cn

**ABSTRACT:** This chapter addresses the increasing vulnerability of coastal regions to typhoons and the consequent power outages, emphasizing the critical role of power transmission systems in disaster resilience. It introduces a framework for assessing and enhancing the resilience of these systems against typhoon impacts. The approach integrates a hybrid-driven model for system failure analysis and resilience assessment, employing both data-driven and model-driven techniques. It includes a unique method to identify system vulnerabilities and optimal strategies for resilience enhancement, considering cost-effectiveness. The efficacy of this method is demonstrated through simulations on the IEEE RTS-79 system under realistic typhoon scenarios, showcasing its potential to guide planners in making informed decisions for disaster resilience.

**Keywords:** Resilience design; power systems; comprehensive failure model; resilience assessment; hybrid data-model driven approach

## 1. Introduction

### 1.1 Background

China ranks among the nations most frequently hit by typhoons globally. Typhoons bring along devastating effects such as severe winds and torrential rains, which significantly threaten the electrical grid in affected regions, leading to substantial economic damage. For instance, in 2016, typhoon "Meranti" made landfall in Xiamen, Fujian, resulting in the failure of 45 substations, 2,837 circuits, impacting 60,301 transformers, and leaving 3.312 million consumers without electricity [1]. Furthermore, in August 2019, the super typhoon "Lekima" hit the southeastern shores of China, leading to over 4,000 circuit failures in Zhejiang among other areas, and disrupting electricity for 67.695 million users [2]. The rising global temperatures have been linked to an increase in the occurrence of such typhoon-induced disruptions to the electrical infrastructure in coastal regions. Additionally, the ongoing incorporation of renewable energy sources into the grid adds further complexity to maintaining power system stability [3-5]. Consequently, pinpointing vulnerable spots within the transmission network and bolstering the grid's robustness efficiently and precisely has become an urgent task to address.

### 1.2 Literature review

In light of the aforementioned challenges, scholars are actively exploring methods to effectively evaluate and enhance the resilience of transmission systems in the face of severe natural disasters. Given the interdisciplinary and complex nature of this subject, various studies are conducted within distinct specializations. This chapter categorizes the relevant research into three main segments: i) Development of fault models, ii) Formulation of resilience metrics, iii) Determination of optimization strategies for transmission systems [6, 7], as depicted in Fig. 1.

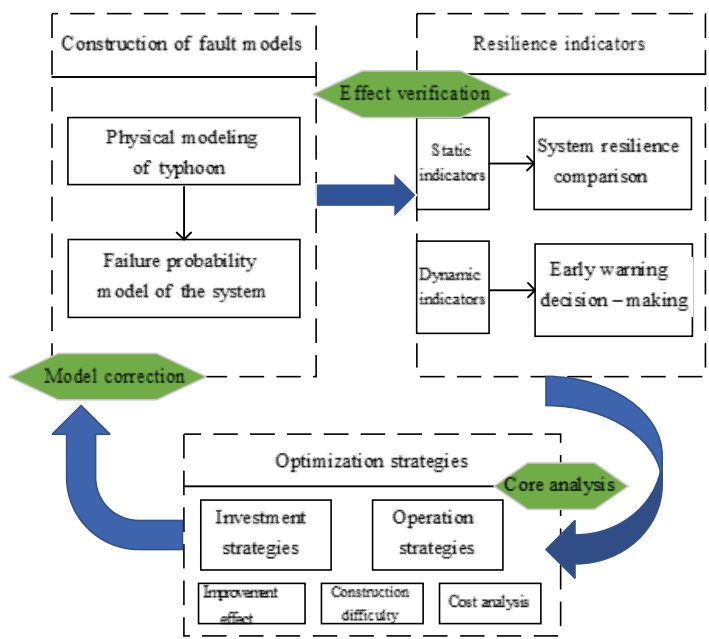

**Fig**. 1 Overview of research areas in power system resilience

### 1.2.1 Construction of fault models

A typhoon represents a multifaceted weather-related calamity capable of triggering a variety of destructive events such as strong winds, heavy rainfall, and flooding. These events significantly increase the risk of failures within electrical transmission systems. To counteract this, the development of a wind field model is crucial for assessing wind velocities across different areas, thereby estimating the failure probabilities of various transmission paths. The pioneering wind field model by Battes, mentioned in Ref. [8], utilized an empirical formula to depict variations in wind speed. Following this model, the research in Ref. [9] determined the overall failure rate of transmission lines through integral theory and formulated a framework for evaluating resilience. Unfortunately, these works mainly focused on wind's impact, neglecting related hazards like heavy rain and the specific topography of the transmission network. In Ref. [10], researchers analyzed historical data from Guangdong's coastal transmission lines to correlate typhoon specifics, terrain features, and failure rates across different voltage levels, subsequently developing a logistic regression model for risk assessment. Further exploration into transmission line failure models by authors in Refs. [11]-[14] applied structural force theories, with Fu et al. in Ref. [11] evaluating the potential for tower collapses and disconnections due to both typhoons and rainstorms. This involved creating a model for the loads induced by these events and examining the dynamic reactions of transmission paths to wind and rain loads. However, due to the high computational demands of the structural force-based models, their application in extensive studies remains challenging. These models also lacked the ability to incorporate certain details, such as micro-topographic and corridor specifics. Later studies, Refs. [15]-[17], employed the Monte Carlo method to simulate the failure probabilities of electrical components during typhoon conditions, using extensive historical data on typhoon characteristics. Meanwhile, research detailed in Refs. [18]-[20] introduced a hybrid data- and model-driven approach to model transmission line failures during typhoon events, but these works did not delve into analyzing or enhancing system resilience.

**1.2.2 Establishment of resilience indicators**

The primary aim of assessing the resilience of power systems is to develop a comprehensive index system and theoretical framework for resilience evaluation. Research to date has identified two main categories of resilience metrics: static and dynamic. Studies cited in Refs. [21, 22] utilized the Monte Carlo simulation to evaluate different scenarios, calculating static reliability indices to gauge the resilience of the investigated systems. Yet, as system reliability is enhanced, sampling outlier states becomes increasingly challenging, which in turn, diminishes the efficiency of the evaluation process. Alternatively, Ref. [23] employs a state enumeration method to calculate the system's dynamic reliability index, offering insights into the system's adaptability in real-time. However, this approach is not feasible for larger systems due to the exponential growth in potential system states. Refs. [24, 25] introduced an innovative approach, the impact-increment-based state enumeration (IISE), transforming conventional enumeration into an impact-increment framework. This method simplifies the computation of state probabilities, significantly enhancing assessment efficiency and precision.

**1.2.3 Optimization strategies**

Optimization strategies are broadly categorized into investment and operational approaches. Investment strategies, as discussed in Refs. [26]-[29], focus on enhancing system resilience through infrastructure redundancy and component fortification, a costly but effective measure. On the other hand, operational strategies, explored in Refs. [30]-[33], involve integrating distributed generation into the system to improve service reliability post-fault, though this can complicate resilience evaluation due to increased grid volatility, as noted in Ref. [34].

Despite extensive research on transmission system resilience against typhoons, existing studies face several challenges:

1) Modeling complexities arise from identifying influential factors on transmission corridor failure rates and integrating the cumulative spatial and temporal impact of typhoons with multifactorial information.

2) Difficulty in achieving optimal evaluation results with different data-driven optimization approaches. Ref. [9] introduced a targeted planning framework to assess resilience, utilizing a hybrid empirical-theoretical model for typhoon wind speed decay. Additionally, leveraging historical transmission system data, a data-driven approach was integrated as a corrective measure. Ref. [19] inspired the use of the Gini index technique, the out-of-bag (OOB) error approach, and the entropy weight method for precise feature weight determination, enabling a thorough identification of system vulnerabilities for more accurate resilience enhancement.

**1.3 Contributions of this chapter**

This chapter advances existing methodologies by integrating both traditional model-driven and modern data-driven approaches to enhance the resilience of power transmission systems against typhoons. Traditional methods often rely on deterministic models which simulate disaster impacts using historical weather data but may not capture real-time variability effectively. In contrast, data-driven methods utilize machine learning to analyze large datasets for predicting potential failures, providing dynamic adaptability absent in traditional models. By synthesizing these methodologies, our hybrid approach leverages the robustness of physical models and the predictive accuracy of data analytics. This integration facilitates a more nuanced assessment of system vulnerabilities and enables the development of more effective resilience strategies. Comparative experiments demonstrate that the hybrid model provides superior predictive capabilities and operational insights, significantly enhancing system resilience over methods that rely solely on traditional or data-driven techniques.

The specific contributions of this study include:

1) This chapter proposes a hybrid data-model driven approach for resilience assessment in electricity transmission systems, blending optimal load reduction strategies with a comprehensive probabilistic failure model. This method not only leverages model-driven insights into the physics of wind disasters and wind load mechanics but also harnesses data-driven analytics for accurate failure rate predictions.

2) Acknowledging the varying judgment principles of different data-driven models, this study presents a multi-attribute decision-making technique known as the analytic hierarchy process-weighted arithmetic averaging (AHP-WAA). This approach combines objective data weight analysis with subjective expertise to identify the most effective data-driven strategies.

3) Through simulation tests on the IEEE RTS-79 test system, located along Guangdong's coastline, the chapter validates the effectiveness and applicability of the proposed methodology, highlighting its practical benefits in enhancing the system resilience in the face of typhoon impacts.

**1.4 Structure of this chapter**

The subsequent sections of this chapter are organized in the following manner: Section 2 outlines the detailed failure probability model of the system, employing a hybrid data-model driven approach. Section 3 concentrates on establishing indicators for evaluating system resilience and identifying enhancement strategies. Section 4 details a case study using the IEEE RTS-79 test system, and Section 5 summarizes the conclusions derived from the study.

**2. Construction of fault model**

Transmission system's resilience encapsulates its capacity to endure disturbances, resist disruptive forces, and swiftly recover operational loads [35]-[37]. Precise quantification of the system's vulnerability to typhoon-induced failures is pivotal for assessing and enhancing system resilience. The failure rate of a transmission system is influenced not only by the wind speed of a typhoon but also by secondary calamities, the local micro-topography, and specific characteristics of the corridors. Conversional model-driven approaches often struggle to simultaneously account for multifaceted disaster scenarios and face challenges in incorporating weak-feature factors into models. Conversely, purely data-driven methods are overly dependent on historical data, which is often insufficient for a comprehensive evaluation.

Given these limitations, this chapter proposes a hybrid approach that primarily relies on model-driven methods, with data-driven elements serving as adjustments. This approach aims to accurately evaluate the impacts of significant meteorological disasters on the system, taking into account additional influencing factors, thereby enhancing the precision of resilience assessments.

**2.1 Model-based failure analysis for transmission corridors**

Typhoon disasters inflict progressive damage on power transmission systems, necessitating a detailed examination of cumulative damage effects. This segment outlines the development of a wind field attenuation model to simulate how typhoon wind speeds affect transmission corridor failure rates.

**2.1.1 Wind field model**

The wind field of a typhoon is characterized by a warm-core structure, marked by low pressure at the center and higher temperatures around it. Given that the fundamental parameters of the typhoon are established, the development of a wind field model typically relies on historical data and theoretical models. This chapter adopts the Batts wind field model, which offers more effective simulation outcomes and enhanced computational efficiency, as depicted in Fig. 2.

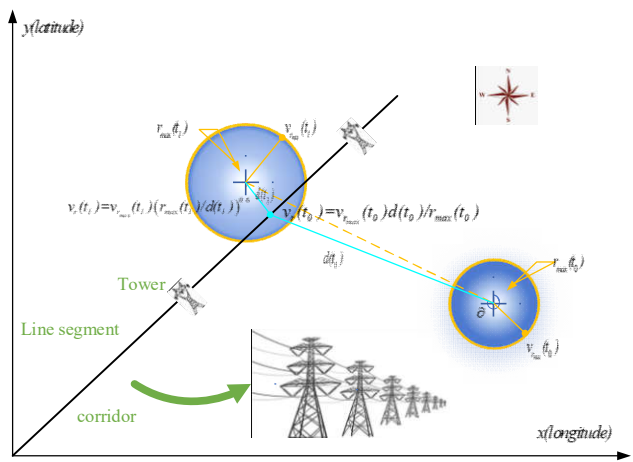

**Fig**. 2 Wind field model of the typhoon

In the model, it is postulated that pressure diminishes over time. The typhoon's pressure decay model is represented as follows:

$$P(t) = \Delta P_0 - [0.02 + 0.02\sin(\partial)]t \tag{1}$$

where $P(t)$ signifies the central pressure difference at moment $t$, $\Delta P_0$ deonts the initial pressure difference, and $\partial$ indicates the angle between the movement direction and north. Subsequently, the variation of peak wind speed over time as the typhoon transits is given by

$$v_{r_{max}}(t) = 0.865K\sqrt{\Delta H(t)} + 0.5v_T \tag{2}$$

where $K$ denotes the model's coefficient [4], and $v_T$ stands for the typhoon's horizontal speed.

The model for the radius of maximum wind speed is derived from [38]:

$$r_{max}(t) = \exp\left(2.63 - 5.086 \times 10^{-5}\Delta H(t)^2 + 0.0395 y(t)\right) \tag{3}$$

where $y(t)$ denotes the latitude of the typhoon center at moment $t$.

The real-time model of wind speed is formulated by [39]:

$$v_r(t) = \begin{cases} v_{r_{max}}(t)d(t)/r_{max}(t), & d(t) \leq r_{max} \\ v_{r_{max}}(t)(r_{max}(t)/d(t))^{0.6}, & d(t) > r_{max} \end{cases} \tag{4}$$

where $d(t)$ denotes the distance from the typhoon center to the corridor at $t$, and $v_{r_{max}}(t)$ represents the maximum wind speed at the corridor.

### 2.1.2 Cumulative failure model of the transmission system

The primary manifestations of typhoon-induced damage to the system are the interruption of lines and the collapse of towers, whereas transformers and cables are generally less susceptible to the impacts of typhoons. Consequently, this chapter focuses on transmission towers and lines, aiming to develop a failure model for transmission corridors during typhoon events. Given the extensive nature of transmission corridors, which can span considerable distances, different segments may be subjected to varying wind speeds. To address this, the study approaches the corridor as a sequential arrangement of multiple tower-line units, where each unit consists of a transmission tower and the connecting line segment. To facilitate analysis, this chapter uses corridor $m$ as a case study.

$$\lambda_{m,l}(t_i) = \exp\left[11 \times \frac{v_{m,l}(t_i)}{v_{d,\text{line}}} - 18\right]\Delta l \tag{5}$$

where $\lambda_{m,l}(t_i)$ and $v_{m,l}(t_i)$ are the failure rate and the wind speed of the $l-th$ section of corridor $m$ at moment $t_i$, $v_{d,\text{line}}$ is the design wind speed of corridor $m$ [9], and $\Delta l$ stands for the $l-th$ section of corridor $m$.

The failure probability model for tower $k$ in corridor $m$ is then derived using the Batts typhoon model.

$$\lambda_{m,k}(t_i) = \begin{cases} 0, & v_{m,k}(t_i) \in [0, v_{d,\text{tower}}] \\ e^{\gamma}[v_{m,k}(t_i) - 2v_{d,\text{tower}}], & v_{m,k}(t_i) \in [v_{d,\text{tower}}, 2v_{d,\text{tower}}] \\ 1, & v_{m,k}(t_i) \in [v_{ex}, \infty] \end{cases} \quad (6)$$

where $\gamma$ denotes the model coefficient, $v_{d,\text{tower}}$ represents the design wind speed of the tower.

In the failure probability model, time is segmented, and wind speed acts as the input. The total duration is divided into intervals to calculate the cumulative failure rate of the transmission corridor. Throughout the typhoon's landfall, the cumulative failure rates for section $l$ and tower $k$ of corridor $m$ are determined by

$$\begin{aligned} p_{m,l} &= 1 - \exp\left(-\int_0^T \lambda_{m,l} dt\right) \\ p_{m,k} &= 1 - \exp\left\{-\int_0^T [\lambda_{m,k}/(1-\lambda_{m,k})]\right\} dt \end{aligned} \quad (7)$$

Each corridor, being a series of widely spaced transmission towers and lines, is considered a unit. If any section fails, the entire corridor is deemed failed. The model for corridor failure probability is detailed in [40, 41]:

$$p_m = 1 - \prod_1^K (1 - p_{m,K}) \prod_1^T (1 - p_{m,T}) \quad (8)$$

where $T$ and $K$ are the counts of line segments and transmission towers in corridor $m$.

### 2.2 Data-driven failure model for transmission corridors

This study calculates the correction coefficients for each tower-line unit in corridors, integrating them with a series formula to determine the corridor's overall failure rate. It starts by weighting key factors influencing failure rates through Gini index, OOB error, and Entropy methods. The AHP-WAA method, guided by expert judgment, selects the optimal scheme and establishes the correction coefficients based on data from the system's tower-line units.

#### 2.2.1 Determination of feature factors

Corridor failure stems from internal factors like design wind speed and tower age, and external forces such as wind and environmental conditions. This chapter combines model-driven and data-driven methods to develop a wind load model and refine failure rate correction coefficients using key internal and external factors. Machine learning techniques adjust for their varying influences, with Fig. 3 showing their impact on failure probability [42].

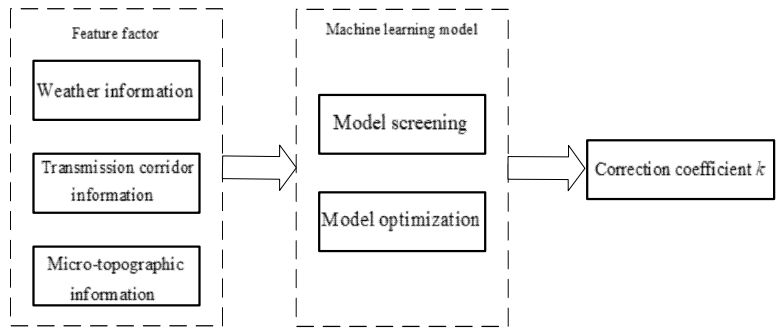

Fig. 3 The calculation process of the correlation coefficient

## 2.2.2 Feature factors weight evaluation schemes

Data-driven techniques have proven to be highly effective in the analysis and optimization of power systems, as highlighted in prior research [43, 44]. These methods, however, differ in their evaluative principles, leading to variations in their outcomes. To accurately determine the significance of various influencing factors in line with real-world conditions, this chapter applies three established data-driven methods in power system evaluations—Random Forest (RF) Gini index, OOB error, and entropy weight. Using these results and expert insights, the AHP-WAA is chosen to determine the optimal correction coefficient scheme.

The Random Forest algorithm stands as a pivotal ensemble learning strategy, leveraging the bagging technique to address classification, regression, and various analytical challenges. It constructs an aggregation of decision trees, where each tree's root node is populated with the entire dataset. Subsequent nodes within each tree are split based on criteria designed to minimize impurity, proceeding until predefined termination conditions are met. This process is detailed in Fig. 4 [45, 46].

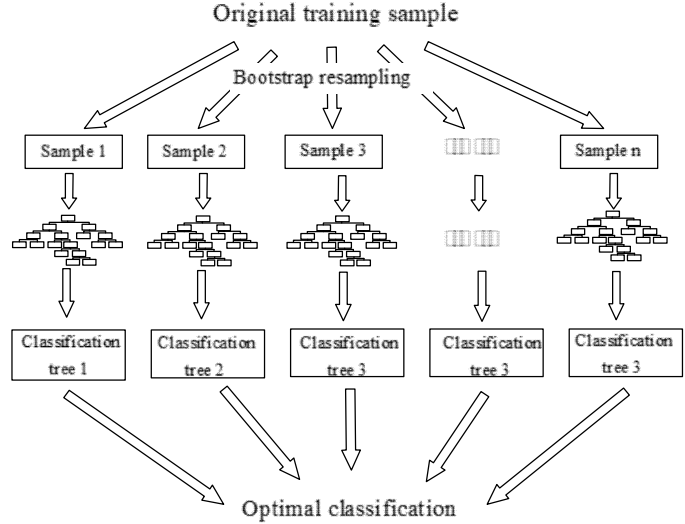

Fig. 4 Random Forest frame diagram

This chapter sets $m$ feature variables $y_1, y_2, \cdots y_m$, and calculates the importance of the Gini method $VIM_j^{gini}$ and the importance of the OOB error method $VIM_j^{oob}$ for the feature variable $y_j$.

1) Gini index method

The Gini index method calculates the average change in node impurity split within the random forest model, as outlined in [47]:

$$GI_m = 1 - \sum_{kI=1}^{|kI|} p_{mkI}^2 \qquad (9)$$

where $KI$ represents the number of categories and is used in this study to differentiate between fault and non-fault states of the transmission corridor, $P_{mKI}$ refers to the proportion of category $KI$ at node $m$.

The importance of each feature variable $y_j$ is determined using the difference in Gini values from the two subsequent nodes, computed as

$$VIM_{jm}^{gini} = GI_m - GI_i - GI_r \tag{10}$$

where $GI_i$ and $GI_r$ represent the Gini values associated with the two nodes located behind the branch..

Given $n$ decision trees in the RF and $y_j$ appearing $R$ times across these, the average importance score is

$$\begin{aligned} VIM_{ij}^{gini} &= \sum_{r=1}^{R} VIM_{jr}^{gini} \\ VIM_{j}^{gini} &= \frac{1}{n}\sum_{i=1}^{n} VIM_{ij}^{gini} \end{aligned} \tag{11}$$

Finally, the importance scores are normalized across all features to assign relative weights:

$$VIM_{j}^{gini} = \frac{VIM_{j}^{gini}}{\sum_{i=1}^{m} VIM_{i}^{gini}} \tag{12}$$

2) OOB error method
The OOB error method assesses the impact of noise on the out-of-bag error rate of each feature in a random forest, where significant changes post-noise introduction indicate a feature's importance. The difference between OOB error rates before and after adding noise serves as the normalized permutation importance of the feature. This approach, while highly accurate and simple, struggles with small and low-dimensional datasets.

The difference between the error rates serves as the normalized permutation importance, which is calculated as

$$VIM_{j}^{OOB} = \frac{\sum_{1}^{n}(e_{j2} - e_{j1})}{n} \tag{13}$$

where $n$ denotes the count of decision trees, $e_{j1}$ and $e_{j2}$ are respectively the OOB error rate of the feature variable $y_j$, and its value after adding noise.

3) Entropy weight method
This method uses the intrinsic information in dataset features to calculate weights objectively, without subjective bias. However, it often overlooks the significance of certain indicators and does not effectively reduce the number of evaluation dimensions, leading to weights that may not align with expected outcomes. It involves normalizing indicators, calculating their entropy, and determining their weights based on the relative entropy values [48, 49].

This method involves normalizing indicators:

$$x_{ij}^{'} = \frac{\max\{x_{1j},\ldots,x_{nj}\} - x_{ij}}{\max\{x_{1j},\ldots,x_{nj}\} - \min\{x_{1j},\ldots,x_{nj}\}} \tag{14}$$

where $j$ denotes the number of indicators, and $i$ represents the sample size of the indicators. Calculate the entropy value of index $j$:

$$e_j = -k \sum_{i=1}^{n} \frac{x_{ij}}{\sum_{i=1}^{n} x_{ij}} \ln\left(\frac{x_{ij}}{\sum_{i=1}^{n} x_{ij}}\right) \tag{15}$$

And finally, determine weights based on entropy:

$$w_j = \frac{1 - e_j}{\sum_{j=1}^{m} 1 - e_j} \tag{16}$$

### 2.2.3 AHP-WAA multi-attribute scheme selection

The three described methods employ objective weighting approaches. In this chapter, the AHP-WAA method, which combines both subjective expertise and objective data analysis, is used for evaluating and prioritizing feature variables, offering a balanced approach to weight determination by incorporating systematic pairwise comparisons and expert judgment, as referenced in [50]-[52]. This strategy ensures a comprehensive assessment that aligns with both empirical evidence and practical insights.

This chapter utilizes $n$ schemes to calculate weights, $w$ feature variables, and builds a decision matrix $Y$ based on the weight results:

$$Y = \begin{bmatrix} y_{11} & y_{12} & \cdots & y_{1w} \\ y_{21} & y_{22} & \cdots & y_{2w} \\ \vdots & \vdots & & \vdots \\ y_{n1} & y_{n2} & \cdots & y_{nw} \end{bmatrix} \tag{17}$$

where $y_{nw}$ denotes the weight of factor $w$ in scheme $n$.

To calculate scheme attribute weights, we first determine the average weight of each feature variable by analyzing the ratio of its total value across $n$ schemes to the total values of all features. We then leverage expert scoring to assess these averages and experience to construct a matrix of relative importance weights $A = (a_{ij})_{w \times w}$, where $a_{ij}$ stands for the ratio of importance ratio of feature $i$ to index $j$. Finally, the matrix $A$ is normalized to drive the relative weight vector $q = [q_1\ q_2\ \cdots\ q_w]^T$. The weight vector $D$ is obtained by

$$D = Y \times q \tag{18}$$

By this means, the optimal scheme is selected based on the maximum value in $D$ as the decision criterion.

### 2.2.4 Calculation of correction coefficient

Based on the optimal evaluation scheme, the correction coefficient for the failure rate is derived from each factor's corresponding weight. The process involves the following steps:

1) Setting the boundary value of the comprehensive score

Differences in feature variables such as rainfall intensity, maximum wind speed, slope, wind angle, and altitude positively influence the transmission corridor's fault correction rate, while operation time and design wind speed have negative impacts [53, 54]. Therefore, positive correlations use a '+' sign, and negative correlations use a '-' sign. The range for each feature variable is chosen according to actual conditions as shown in Table 1, from which the boundary value for the comprehensive score is determined by

$$W_{B_{max}} = \max\left(\sum_{i=1}^{m}\omega_i x_{B_{max},i}, \sum_{i=1}^{m}\omega_i x_{B_{min},i}\right)$$
$$W_{B_{min}} = \min\left(\sum_{i=1}^{m}\omega_i x_{B_{max},i}, \sum_{i=1}^{m}\omega_i x_{B_{min},i}\right) \tag{19}$$

where $\omega_i$ denotes the weight of the feature variable $i$, $m$ refers to the total count of feature variables, and $x_{B_{max},i}$, $x_{B_{min},i}$ are the boundary quantity set by the feature variable $i$.

**Table** 1 Variable value range

| Maximum wind speed (m/s) | Design wind (m/s) | Rainfall intensity (mm/h) | Wind angle (°) | Altitude (m) | Operation time (year) | Slope (°) |
|---|---|---|---|---|---|---|
| 0~60 | 20~50 | 0~60 | 0~180 | -20~150 | 0~40 | 0~180 |

2) Calculate the composite score

$$W = \omega x^T = \sum_{i=1}^{m}\omega_i x_i \tag{20}$$

3) Determine the correction coefficient

This work scales the revision coefficient value to the range [0.9-1.4] based on reference [19]:

$$k = 0.5\frac{W - W_{B_{min}}}{W_{B_{max}} - W_{B_{min}}} + 0.9 \tag{21}$$

## 2.3 Determination of the comprehensive failure rate

To optimize computational efficiency and account for the minimal variation in feature factors over specific distances, this work employs a grid method, partitioning the studied region into 1 km by 1 km segments. This section examines the impact of wind speed on transmission corridors during typhoons, assessing the corridor's failure rate with a model-driven method based on key factors: design wind speed, operational history, maximum wind speed, rainfall, slope, wind direction, and altitude. A correction coefficient is then calculated through a data-driven approach, streamlining the resilience evaluation process. The comprehensive failure rate of corridor $m$ can be formulated by

$$p_{mc,i} = k_{m,i} p_{m,i} \quad i = 1, 2, \cdots, n$$
$$P_{mc} = 1 - \prod_{i=1}^{n}(1 - p_{mc,i}) \quad (22)$$

Here $k_{m,i}$ denotes the correction coefficient for each tower-line analysis unit $i$ in corridor $m$, $p_{m,i}$ stands for the cumulative failure rate of that unit, $p_{m,i}$ also refers to the comprehensive failure rate of the unit, $n$ represents the count of such tower-line analysis units in corridor $m$. Fig. 5 depicts the construction process of the comprehensive failure probability model for corridor $m$.

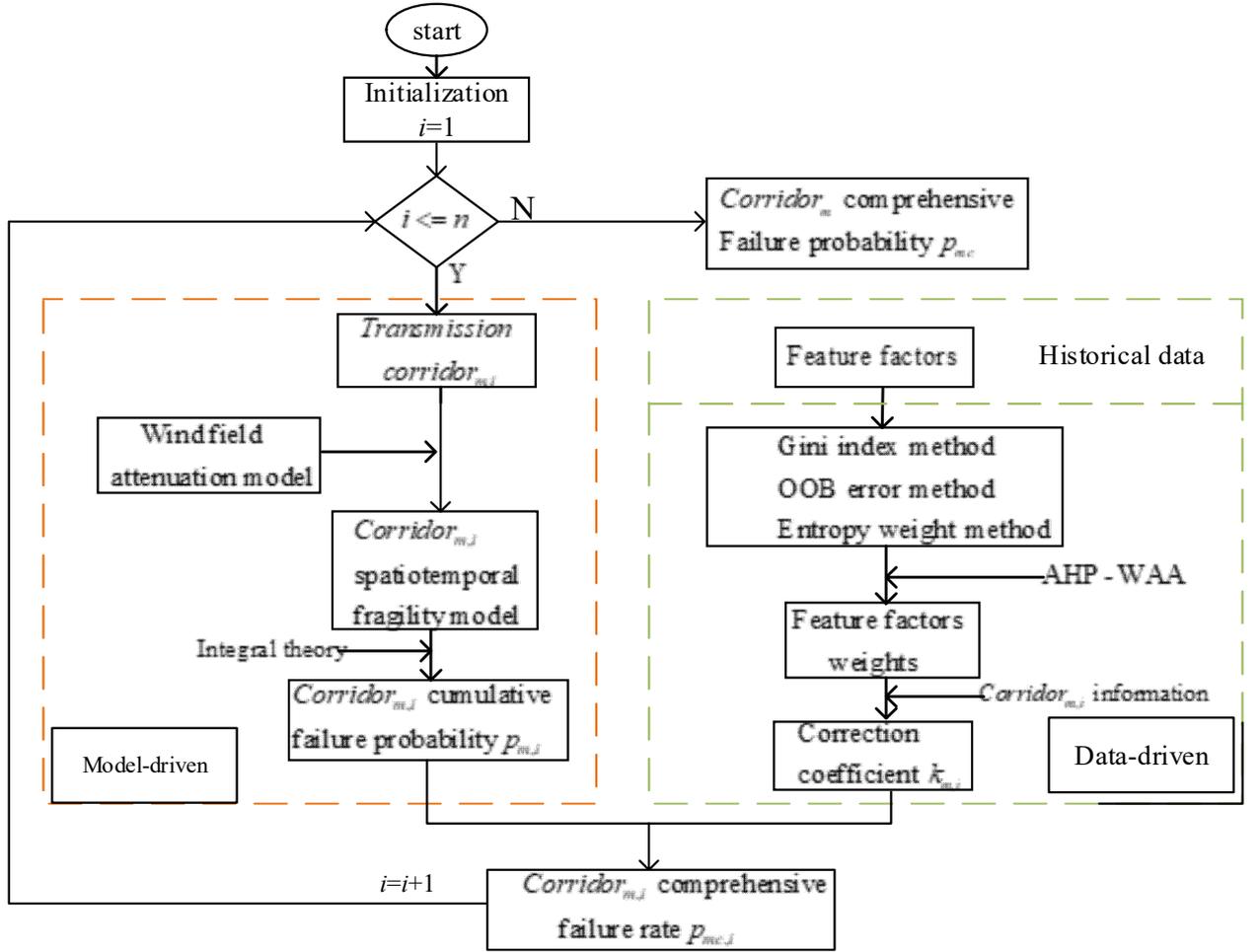

**Fig**. 5 Comprehensive failure probability model for corridor $m$

Similarly, by gathering data from other corridors, we traverse the correction coefficients for their tower-line units to construct the system's comprehensive failure probability model.

## 3. Resilience assessment and improvement methods

Creating resilience assessment indicators is crucial for quantifying the resilience of a system and pinpointing its vulnerabilities. By evaluating system-level indicators, we can quantify the extent of performance degradation resulting from typhoons, ensuring it aligns with planned expectations. Additionally, by examining specific corridor indicators, we can pinpoint the system's vulnerable corridors. This enables the implementation of focused optimization strategies to enhance the system's overall resilience.

### 3.1 Analysis of typhoon probability model

The research utilizes a state enumeration method to analyze typhoon occurrence probabilities, integrating empirical distributions of key typhoon parameters such as moving speed $v_T$, central pressure difference $\Delta H_{0,w}$, and direction $\theta$, which adhere to log-normal and bi-normal distributions respectively [39]. These parameters are combined in a probability model to estimate the occurrence of typhoon $w$ using the formula:

$$P_w = P_r(\Delta H_{0,w}) P_r(v_{T,w}) P_r(\theta_w) \tag{23}$$

Here $P_r$ is the occurrence probability. By doing so, a typhoon wind field model can be obtained for any location within the affected area, thereby providing a data foundation for constructing resilience indices of the transmission system during a typhoon.

### 3.2 The system-level resilience indicator

System-level resilience metrics are vital for measuring transmission system resilience. As depicted in Fig. 6, the resilience triangle, a common method in current literature, quantifies system resilience by integrating performance reduction over time. A higher index indicates lower system resilience [55].

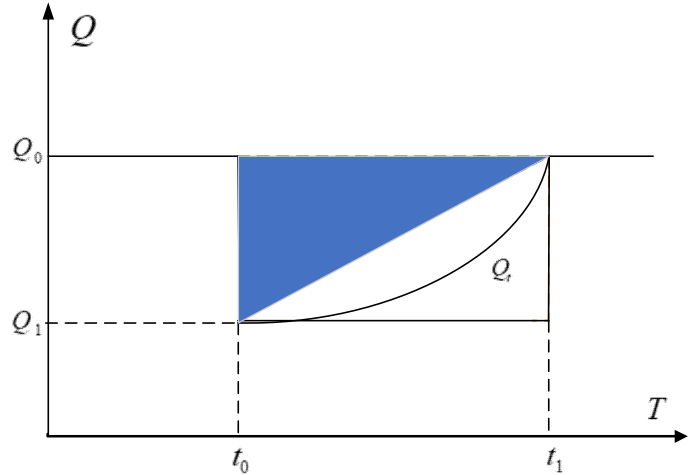

**Fig**. 6 Resilient triangle and rectangle

For a typhoon, $Q_0$ denotes the system's normal operational state, $Q_1$ represents the state at peak distress, $Q_t$ stands for the performance curve. The resilience index is calculated by integrating the performance reduction over time, expressed as

$$\Upsilon = \int_{t_0}^{t_1} [Q_0 - Q(t)] \mathrm{d}t \tag{24}$$

Traditional metrics fail to account for typhoon unpredictability and intensity variations, making uniform measurements imprecise. To address this, the resilience triangle is modified to a rectangle, simplifying the resilience index formula to

$$R = E[\Upsilon] \approx E[Q_0 - Q_1] = \sum_{w \in W} P_w (Q_0 - Q_1)_w \tag{25}$$

where $E$ denotes the expected value of the minimum system load reduction, $P_w$ represents the probability of the typhoon occurrence, and $W$ denotes potential typhoons.

The IISE method is applied to compute and mitigate the impact of severe faults, redefining the resilience index as

$$R_{sys} = \sum_{w=1}^{W} P_w \sum_{j=1}^{J} \sum_{s \in \Omega_j} \left( \prod_{i \in s} p_{w,i} \right) \Delta I_{w,s} \tag{26}$$

where $\Omega$ indicates the set order of fault events, $s$ is the count of faulted corridors, $J$ is the highest failure order, $p_{w,i}$ is the failure rate of corridor $i$ against typhoon $w$, and $\Delta I_{w,s}$ quantifies the incremental influence of state $s$ during the typhoon.

### 3.3 The corridor-level resilience index

In this study, we use the corridor-level resilience index $R_m$ to assess corridor fragility within the power system, defined as

$$R_m = R_{sys} - R_{sys.m} \tag{27}$$

where $R_{sys}$ is the resilience index the system without faults and $R_{sys.m}$ denotes the index when corridor $m$ experiences a fault during a typhoon. This index measures the impact of a corridor's failure on overall system resilience, indicating that higher values of $R_m$ signify more severe consequences for the system, thereby highlighting the importance of strengthening corridor $m$ to enhance system recovery [56, 57].

### 3.4 Strategies for improving resilience

Enhancing transmission corridors typically involves upgrading system components or incorporating additional redundancies. Upgrades might include replacing less resilient structures with those offering greater resistance to disasters, such as substituting traditional corner towers with cat-head type towers for improved performance. Adding backup corridors in areas prone to high failure rates can also decrease the system's overall failure likelihood. However, the cost of such enhancements varies due to differences in corridor lengths and geographical conditions. Thus, resilience strategies should be developed based on reinforcement costs, effectiveness, and construction challenges, allowing for a comparison of strategy benefits to identify the most effective reinforcement approach.

## 3.5 Resilience assessment and improvement process

Fig. 7 illustrates a planning-targeted resilience assessment method that begins by analyzing key typhoon parameters to predict occurrences. It then constructs a comprehensive failure probability model based on disaster information and uses MATPOWER within power flow constraints to calculate optimal load reductions for various disaster scenarios. The process includes developing a system-level resilience index using the IISE method; if this meets a set threshold, the system's optimal plan is implemented. If not, the process evaluates corridor-level resilience to identify and strategize on system weaknesses, ultimately selecting a resilience improvement strategy based on cost-effectiveness and construction feasibility.

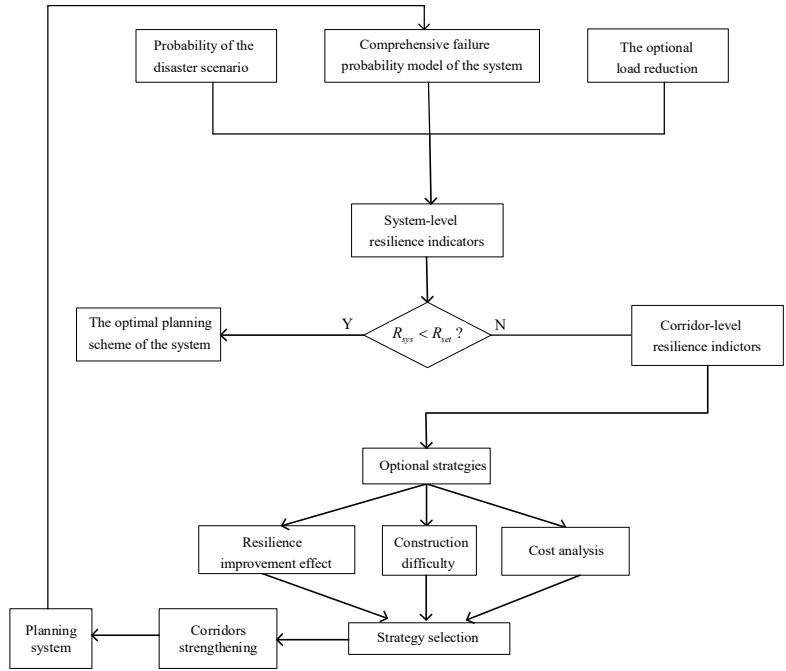

**Fig**. 7 Resilience assessment and improvement flowchart

## 4. Case study

The IEEE RTS-79 test system serves as the basis for assessing the proposed method's efficacy and practicality [58]. Given Guangdong Province's vulnerability to typhoons, the system is mapped onto this coastal region for analysis ease. This chapter utilizes real data from typhoon "Mangkhut" for weather analysis. Transmission towers are spaced 500 meters apart, with each segment between towers also measuring 500 meters, incorporating realistic geographic locations for each tower-line unit [59]. Furthermore, a comparative analysis is introduced by including a control test that calculates the corridor failure rate using a model-driven approach with multifactorial considerations, illustrating the proposed method's superiority [60]-[62]. The layout of the IEEE RTS-79 system and the corresponding topographical features of Guangdong's coast are depicted in Fig. 8, showing the exact locations of the transmission corridors.

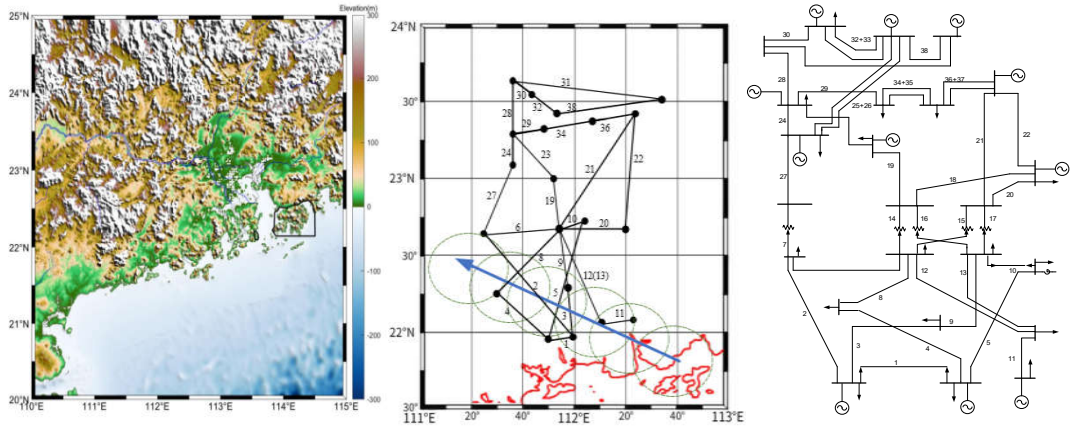

a) Wiring schematic diagram     b) Topographic diagram     c) Location diagram

**Fig**. 8 Wiring principle and topographic

Typhoon "Mangkhut" made landfall at coordinates 21.8°N and 112.7°E, with an initial pressure differential of 58 hPa and an average progression speed of 30 km/h. Illustrated in Fig. 8(b), the typhoon's trajectory is indicated by a blue arrow, its wind speed radius by green, and Guangdong Province's coastline by a red curve [9]. For analytical convenience, transmission corridors 3, 4, 8, 11, and 27 have been chosen as representative samples for testing. Fig. 9 illustrates the wind velocity at the center of the aforementioned five corridors.

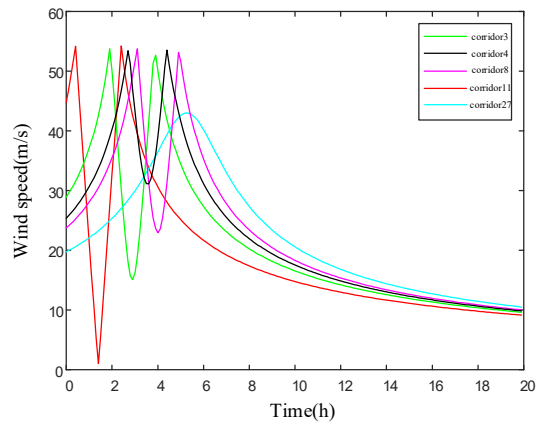

**Fig**. 9 Wind velocity of the transmission corridors

Focusing on corridor 3 as a case study, as corridor 3 moves closer to or further from the typhoon center beyond the wind circle radius, the wind speed increases, then decreases once inside the radius, and diminishes to zero once it again exceeds the radius.

**4.1 Model-driven failure rate calculation**

This section discusses the model-driven failure rate calculation for the IEEE RTS-79 test system, analyzing physical models and calculating failure rates for each grid's transmission corridors, represented by 1km² areas of the same voltage level to minimize computational costs. As depicted in Fig. 10, the cumulative failure rate in the corridors escalates over time, intensifying with higher wind speeds. This methodology applies to calculating the cumulative failure rates for other corridors similarly.

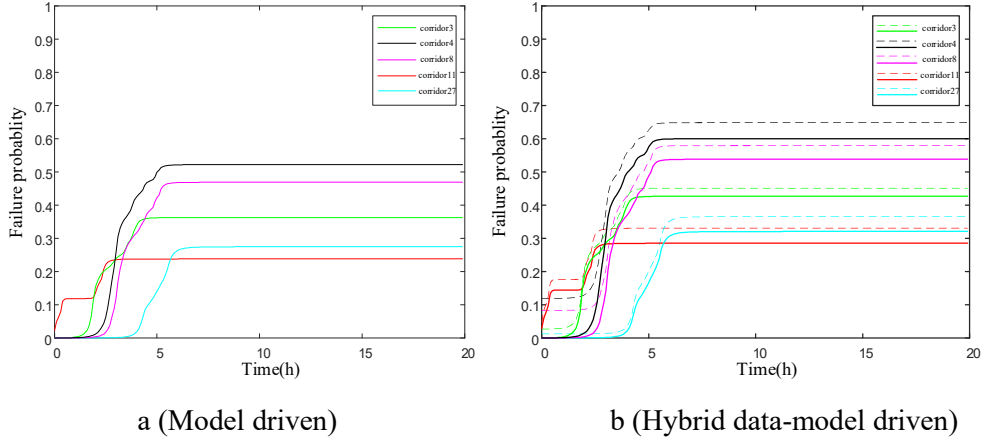

a (Model driven)            b (Hybrid data-model driven)

**Fig**. 10 Failure rates of the transmission corridors

### 4.2 Data-driven failure rate calculation

This research compiles 640 balanced sample datasets from Guangdong Province's transmission corridors affected by typhoons, covering seven key features: maximum wind speed, 24-hour total rainfall intensity, altitude, slope, wind direction, design wind speed, and operational duration. The use of 24-hour rainfall data from the China Meteorological Administration poses challenges for direct analysis of the corridors' responses. Reference [59] introduced methods for converting total rainfall over 24 hours into a 10-minute rainfall intensity measure as follows:

$$R_{10\min} = 27.08 R_{24h}^{0.6021} \tag{28}$$

where $R_{24h}$ and $R_{10\min}$ represent the total rainfall over 24 hours and the intensity of rainfall over a 10-minute period, respectively.

Utilizing historical data, this chapter employs the Gini index, OOB error, and entropy weight methods to sequentially develop three schemes, calculating the importance weights of various variables as depicted in Table 2.

**Table 2** Results of evaluating the importance weights of feature variables

| Schemes | Weights | | | | | | |
|---|---|---|---|---|---|---|---|
| | Max wind speed | Rainfall intensity | Altitude | Slope | Wind angle | Design wind | Operation time |
| Scheme 1 (Gini index method) | 0.236 | 0.146 | 0.182 | 0.078 | 0.105 | 0.081 | 0.169 |
| Scheme 2 (OOB error method) | 0.208 | 0.214 | 0.104 | 0.099 | 0.166 | 0.119 | 0.095 |
| Scheme 3 (Entropy weight method) | 0.158 | 0.139 | 0.186 | 0.130 | 0.129 | 0.136 | 0.122 |

Scheme 1 highlights maximum wind speed at 23.6% and slope at 7.8%; Scheme 2 features rainfall intensity at 21.4% and operation time at 9.5%; Scheme 3 prioritizes altitude at 18.6% and slope at 12.2%, as per Table 2. These variations arise from the distinct methodologies used. The AHP-WAA method integrates these evaluations, with Fig. 11 illustrating average proportions and expert assessments ranking influence on transmission corridors as: maximum wind speed > rainfall intensity > altitude > operation time = wind angle > design wind > slope.

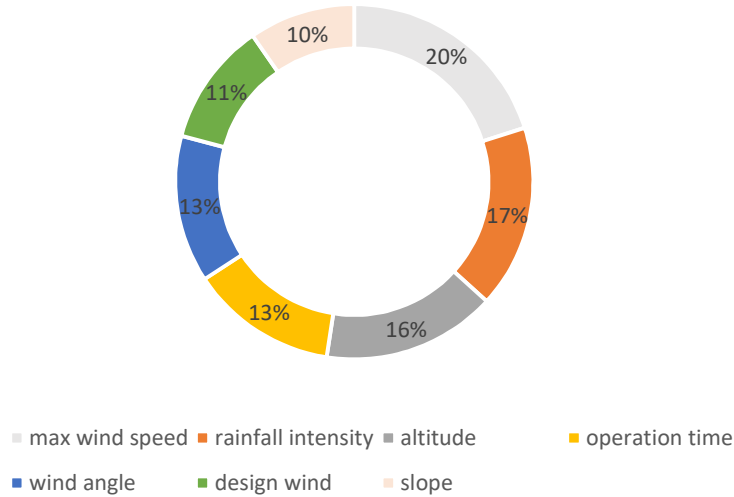

**Fig**. 11 Proportional distribution of different features

Referring to the Santy scale method, which quantifies typhoon impacts like wind speed and rainfall intensity based on historical data, we obtain and rank the relative significance of feature variables. This method enhances our model's precision in assessing transmission corridor vulnerabilities, as shown in Table 3, where the Santy scale's integration with other data-driven techniques illustrates how various typhoon characteristics affect system resilience.

**Table 3** Comparison of levels of different feature variables

| Comparison | Maximum wind speed | Rainfall intensity | Altitude | Operation time | Wind angle | Design wind | Slope |
|---|---|---|---|---|---|---|---|
| Maximum wind speed | 1 | 3 | 4 | 9 | 5 | 7 | 5 |
| Rainfall intensity | 1/3 | 1 | 2 | 6 | 3 | 5 | 3 |
| Altitude | 1/4 | 1/2 | 1 | 5 | 2 | 4 | 2 |
| Operation time | 1/9 | 1/6 | 1/5 | 1 | 1/4 | 1/2 | 1/4 |
| Wind angle | 1/5 | 1/3 | 1/2 | 4 | 1 | 3 | 1 |
| Design wind | 1/7 | 1/5 | 1/4 | 2 | 1/3 | 1 | 1/3 |
| Slope | 1/5 | 1/3 | 1/2 | 4 | 1 | 3 | 1 |

Next, decision scores for each scheme are computed using the AHP-WAA decision-making method, with the results displayed in Table 4.

**Table 4** Scheme scores

| Schemes | Scheme 1 | Scheme 2 | Scheme 3 |
|---|---|---|---|
| Weights | 0.1822 | 0.1753 | 0.1511 |

In Table 4, Scheme 1 scores the highest, aligning closely with subjective human evaluations. Table 2 shows that maximum wind speed, altitude, and operation time significantly impact failure probabilities, whereas the effect of slope is less pronounced. Fig. 10(b) uses a dashed line to show failure rates determined by a model-driven approach that includes rainfall intensity and other variables [60]-[62]. In contrast, the solid line in the same figure represents the total failure rate using the hybrid model discussed in this study. This model's failure rate aligns closely with that seen in the control test in Fig. 10(a). Furthermore, the hybrid model's cumulative failure rate exceeds that of the purely model-driven approach, highlighting an increased risk of failure after adjustments. This arises from the model-driven approach's oversight of multiple factors' influence on corridors, as it optimizes other variables to ideal conditions without considering their combined effects.

**4.3 Resilience assessment and enhancement results**

**4.3.1 System resilience index comparison: hybrid vs. model-driven approaches**

In this section, we compare traditional model-driven methods with innovative hybrid data-model driven approaches for evaluating power transmission systems' resilience against typhoons, using the IISE to calculate the resilience index $R_{sys}$. For example, in Guangdong's coast, where typhoons often disrupt the transmission system, the default failure enumeration is set to 2, highlighting the need for a more detailed vulnerability analysis at the corridor level. We calculate the comprehensive failure rate and the incremental impact of each fault scenario to derive corridor-specific resilience indices according to Eq. (27). Before and after adjustments, we assess the system-level and corridor-level resilience indices for the five most vulnerable corridors, with results presented in Table 5.

**Table 5** Comparison of system resilience indices using different approaches

| Model driven approach | | Hybrid data-model driven approach | |
|---|---|---|---|
| Resilience indicator | Result (MW) | Resilience indicator | Result (MW) |
| $R_{sys}$ | $19.98 \times 10^{-5}$ | $R_{sys}$ | $23.67 \times 10^{-5}$ |
| $R_{27}$ | $8.22 \times 10^{-5}$ | $R_{27}$ | $9.19 \times 10^{-5}$ |
| $R_{11}$ | $6.03 \times 10^{-5}$ | $R_{11}$ | $6.57 \times 10^{-5}$ |
| $R_{13}$ | $3.07 \times 10^{-5}$ | $R_4$ | $4.07 \times 10^{-5}$ |
| $R_{12}$ | $2.94 \times 10^{-5}$ | $R_8$ | $4.06 \times 10^{-5}$ |
| $R_4$ | $2.73 \times 10^{-5}$ | $R_{13}$ | $3.71 \times 10^{-5}$ |

Table 5 indicates that the resilience index of each transmission corridor are comparable in the model-driven approach due to the probability of typhoon occurrence being uniformly calculated across all scenarios. However, the resilience index values of the hybrid-driven approach are higher, indicating a more detailed assessment incorporating multi-factor information which results in a relatively higher cumulative failure rate for each transmission corridor. Notably, the priority order of transmission corridors shifts with the inclusion of correction coefficients, elevating the priority of corridor 4 and corridor 8 over corridor 13. This adjustment stems from the close relationship between corridor-level resilience index values and corridor failure rates. As depicted in Fig. 8(b), corridors 4 and 8 exhibit significantly higher altitudes compared to corridor 13. Furthermore, according to a report by the National Meteorological Centre dated September 16, 2018, corridor 4 was subject to an extreme rainstorm, corridor 8 was located in an area of significant rainfall, and corridor 13 faced heavy rains. As a result, the environmental conditions impacting corridors 4 and 8 were more severe compared to corridor 13. However, when incorporating multifactorial data, the resilience indices of corridors 4 and 8 show marked improvement after adjustments, leading to a shift in corridor priority. Based on these facts, it can be concluded that the traditional model-driven approach estimated resilience indexes significantly lower than those observed, suggesting a lower risk of failure. In contrast, the hybrid approach yielded higher resilience indexes that closely matched the actual damages reported, demonstrating superior real-time adaptability and accuracy.

**4.3.2 Resilience Improvement Effects**

This study examines the integration of redundancy in transmission corridors as a method to enhance the resilience of the transmission system, noting substantial variations in the strengthening costs among different transmission corridors. It's important to highlight that this chapter does not account for construction difficulty in its analysis. Referencing [53, 54], this chapter establishes the cost of the transmission corridor at 1 million dollars per kilometer. Additionally, planners must thoroughly consider the cost-effectiveness of enhancing corridor strategies. With $R_{set}$ defined at $17 \times 10^{-5}$ MW, this chapter outlines six strengthening strategies:

Strategy 1 targets corridor 27, the most vulnerable; Strategy 2 targets corridors 8 and 11; Strategy 3 bolsters corridors 4 and 11; Strategy 4 reinforces corridors 3, 8, and 13; Strategy 5 fortifies corridors 3, 10, and 13; and Strategy 6 strengthens corridors 8, 10, and 13.

Table 6 details the resilience improvements, costs for each strategy, and the cost-effectiveness ratios. Here, 'C' represents the cost of enhancement, 'RE' indicates the decrease in the system-level resilience index, 'ΔRE' is the reduction percentage, and 'C/ΔRE' shows the cost per percentage reduction in the resilience index.

**Table 6** Comparison of resilience improvement effects across different strategies

| Priority | Strategies | C ($) | RE (MW) | ΔRE (%) | C/ΔRE ($) |
|---|---|---|---|---|---|
| 1 | Strategy 1 | $579.63 \times 10^5$ | $9.19 \times 10^{-5}$ | 38.83 | $14.93 \times 10^5$ |
| 2 | Strategy 2 | $692.35 \times 10^5$ | $10.63 \times 10^{-5}$ | 44.92 | $15.41 \times 10^5$ |
| 3 | Strategy 3 | $788.92 \times 10^5$ | $10.64 \times 10^{-5}$ | 45.11 | $17.49 \times 10^5$ |
| 4 | Strategy 5 | $1304.11 \times 10^5$ | $9.11 \times 10^{-5}$ | 38.52 | $33.86 \times 10^5$ |
| 5 | Strategy 4 | $1481.20 \times 10^5$ | $9.77 \times 10^{-5}$ | 41.30 | $35.86 \times 10^5$ |
| 6 | Strategy 6 | $1384.62 \times 10^5$ | $7.06 \times 10^{-5}$ | 30.02 | $46.12 \times 10^5$ |

Table 6 reveals that Strategy 1 offers the best resilience enhancement and the lowest cost-effectiveness ratio at $\$14.93 \times 10^5$, while Strategy 6, though the least effective, has a high cost-effectiveness ratio of $\$46.12 \times 10^5$. If constructing Strategy 1 proves challenging, Strategies 2 and 3 are viable alternatives, both achieving a system-resilience index of $17 \times 10^{-5}$ MW but at a slightly higher cost. Analysis suggests that Strategy 1 is the most advantageous.

Fig. 12 displays the normalized values of cost, resilience enhancement in MW, resilience improvement percentage, and cost-effectiveness for each strategy, scaled to 100% based on the maximum value of each metric. Each strategy is represented by a group of bars, with different colors indicating different metrics: purple for cost, green for resilience enhancement (MW), yellow for resilience improvement (%), and blue for cost-effectiveness.

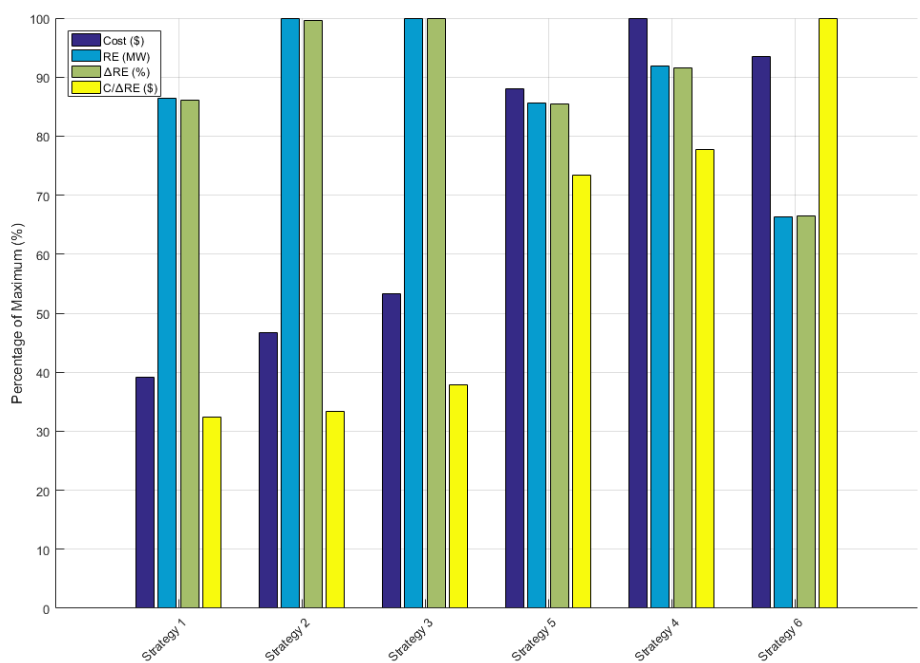

**Fig**. 12 Normalized bar chart of all metrics for strengthening strategies

As shown in Fig. 12, Strategy 1 is highlighted as providing the highest resilience improvement at the most cost-effective ratio, whereas Strategy 6, despite having the lowest improvement percentage, incurs the highest cost per percentage point of resilience gained. The normalization allows for direct comparisons across the strategies, highlighting their relative strengths and efficiencies in a visually engaging manner. This chart not only demonstrates the diverse impacts of each strategy but also aligns with our efforts to enhance the intuitive analysis and presentation of our case study results.

## 5. Conclusions

At present, the accurate evaluation of transmission systems' resilience to typhoon disasters is complicated by uncertainties in typhoon data, transmission corridor information, and micro-topographic specifics. To address these challenges, this chapter presents a planning-focused resilience assessment framework that takes into account the effects of various factors on system resilience in the face of typhoon disasters. The main conclusions are as follows:

1) This chapter utilizes a hybrid-driven resilience assessment framework that integrates both data-driven and model-driven approaches to evaluate the system resilience in the face of typhoon disasters. This comprehensive framework accounts for various uncertain factors and effectively addresses challenges in modeling weak influencing factors like details and corridor data.

2) Regarding the solution approach, the AHP-WAA decision-making method is proposed to accurately assess the weighting for various data-driven schemes. This method combines subjective assessments with objective optimization to yield optimal data-driven evaluation results, effectively mitigating significant discrepancies in evaluation outcomes caused by different data-driven models.

3) The experimental results using the IEEE RTS-79 system validate the effectiveness of the

presented method, demonstrating that the hybrid-driven model outperforms the traditional driven model in enhancing system resilience. This approach accurately identifies transmission system vulnerabilities, aiding decision-makers in devising suitable resilience enhancement strategies.

Future work will concentrate on optimizing the selection of feature factors and sample data, aiming to enhance prediction accuracy and enhance the system resilience. By refining and expanding the feature sample data via data augmentation [63], key factors can be effectively identified, thereby enhancing resilience. Furthermore, real-world transmission systems will be employed to validate the proposed method under applicable experimental conditions. Additionally, extending this resilience assessment approach to evaluate both integrated energy systems [64] and power systems under cyber-attacks [65] presents a promising direction for future research. Moreover, exploring the impact of terrain on construction difficulty will enable the provision of comprehensive resilience improvement strategies for decision-makers in the future.